# Neon ion Irradiation studies on MgB$_2$ Superconductor

A.Talapatra, S.K.Bandyopadhyay[*], Pintu Sen and P.Barat

Variable Energy Cyclotron Centre,

1/AF, Bidhan Nagar, Kolkata-700 064

**Abstract:**

160 MeV Neon ion irradiation has been carried out on MgB$_2$ polycrystalline pellets at various doses. There has not been any significant change in Tc except at the highest dose of $1 \times 10^{15}$ ions/cm$^2$. Increase in resistivity has been noticed. Resistivity data have been fitted with Bloch-Grüneisen function and the values of Debye temperature, residual resistivity and temperature coefficient of resistivity have been extracted for irradiated as well as unirradiated samples. The increase in the resistivity of irradiated samples has been explained in the light of damage in the 3D π bonding network of Boron.

PACS No. 74.70.Ad, 61.72Ji

Keywords: MgB$_2$, Ion irradiation, π-bonding, Coupling constant

[*] Corresponding author. E-mail address: skband@veccal.ernet.in

## 1. Introduction:

The discovery of superconductivity in Magnesium Diboride, $MgB_2$ [1] has attracted a lot of interest because it is a simple binary intermetallic system with fairly high $T_c$. Its conductivity is also quite high. The relatively high $T_c$ and high conductivity place it among potential attractive materials for application in devices. $MgB_2$ is within the clean limit with large coherence length (~5 nm) as compared to HTSC cuprates. The flow of supercurrents in $MgB_2$ is not significantly hindered by grain boundaries [2] which is dominant in cuprates. The observation of significant B isotope effect on superconducting transition temperature $T_c$ [3] and heat capacity measurement [4] indicate that $MgB_2$ is a phonon mediated superconductor [5].

Experimental results suggest that it is a multiband superconductor [6]. $MgB_2$ is built up on hexagonal structure where Magnesium (Mg) layers are sandwiched between two Boron (B) layers. There is a strong $sp^2$ hybrid σ bonding within in-plane B atoms that gives rise to 2D σ bands. The out-of-plane B atoms are coupled by $p_z$ orbitals through Mg atoms giving rise to 3D metallic π bands.

Some particle irradiation studies have been pursued on this system [7-9]. $MgB_2$ behaves quite differently from the HTSC cuprates with respect to particle irradiation. In case of cuprates, the radiation causes knock-out of non stoichiometric oxygen with low binding energy [10]. This brings about a drastic change in superconductivity [11]. The irradiation studies with heavy ions on thin films [8] and protons on bulk materials [7] of $MgB_2$ have not reflected any significant changes in $T_c$ and other superconducting properties. Heavy ions with large deposition energy and high values of displacements per atom (dpa) are expected to bring about changes in bulk samples. In this paper, we are reporting studies of irradiation with Neon ions on bulk pellets of $MgB_2$. We have carried out $T_c$ and resistivity studies with the purpose of investigating irradiation effects on electron phonon coupling (EPC) which is the dominant parameter for controlling superconductivity in this system.

## 2. Experimental:

MgB$_2$ pellets have been synthesized starting from Mg (> 97% pure) and B (99.9% pure) powder with 2 atomic percent Mg in excess to stoichiometry. The volatile Mg was employed in excess to compensate for the loss. The mixture was pelletised by applying uniaxial pressure of 7.5 tons/cm$^2$. The pellets were wrapped in Tantalum foil, put in quartz tube and evacuated to $10^{-2}$ torr. They were purged with 99.9% pure helium quite a number of times to expel the oxygen in air and sealed in helium environment (at ~ 800 torr). The inert gas environment was needed to avoid the formation of MgO phase. The samples were heated in quartz tube at 800$^0$C for two hours and then at 900$^0$C for one hour and finally quenched at 625$^0$C to arrest the phase of MgB$_2$. Formation of MgB$_2$ is exothermic as observed by our Thermogravimetry (TG) and Differential Thermal Analysis (DTA) [12] and hence the employment of tantalum foil as heat sink was essential. The samples thus obtained were characterized by X-ray Diffraction (XRD) pattern taken with Philips PW1710 diffractometer with Cu K$_\alpha$ of wavelength 1.54 Å. Resistivities of the samples as function of temperature were measured in close cycle helium refrigerator (made of Cryoindustries of America) using four probe technique with HP 34220A Nanovoltmeter with resolution of 0.1 nanovolt and Keithley 224 programmable current source. 1mA current was employed. Tc of the unirradiated sample was 38.7K with a transition width ($\Delta T_c$) of ~1.0K.

Irradiation has been carried out with 160 MeV Neon ions available at Variable Energy Cyclotron Centre, Kolkata, at doses of $1\times10^{13}$, $1\times10^{14}$ and $1\times10^{15}$ ions/cm$^2$. The beam current was kept around 100nA to avoid heating. Irradiation was carried out from both sides of the samples to attain maximum bulk damage. Irradiated samples were characterised by XRD and Tc like the unirradiated sample.

## 3. Results and Discussions:

Fig.1 (a-d) shows the XRD patterns of irradiated as well as unirradiated samples. There is no indication of development of any phase other than $MgB_2$ in all the samples. The lattice parameters were obtained from Rietveld analysis using LS1 programme. The values obtained, a=3.09 Å and c= 3.49 Å, are in good agreement with the literature values [1]. The pattern of the sample irradiated at the highest dose shows line broadening corresponding to some characteristic reflection planes (<101> and <110>).

The plots of resistivity versus temperature for all the four samples are shown in Fig. 2. We observe that there is no significant change in $T_c$ indicative of rather insensitivity of $MgB_2$ towards particle irradiation. There is slight decrease in $T_c$ for the sample with the highest dose. The values of $T_c$ and $\rho_{300}$ (room temperature resistivity) are listed in Table 1. There is almost no increase in $\Delta T_c$ excepting at the highest dose. Room temperature resistivity ($\rho_{300}$) of the polycrystalline samples increased with dose except for the lowest dose. Decrease in resistivity for the sample irradiated with the dose of $1x10^{13}$ Neon/cm$^2$ may be due to thermal annealing of the defects, which were initially present in the sintered sample. Similar phenomenon was observed in case of $YBa_2Cu_3O_7$ irradiated at low dose of heavy ions like B [13]. At low dose of irradiation, mobile defects are also seen to increase the long-range ordering in partly ordered metallic alloys [14].

Depth of 160 MeV Neon ion implantation is 106μ, as obtained from Monte Carlo simulation using the code SRIM 2003 [15]. Displacement energy of both Mg and B has been 25eV with lattice binding energy of 3eV [15]. The high binding energy of B is an outcome of strong sp$^2$ hybrid σ bonding between in-plane B atoms. The number of displacements/ion is 2734. The dpa in the range of 106μ obtained thereby is $8.2 \times 10^{-18}$/ion/cm². Energy loss here is larger by a factor of $10^2$ as compared to that caused by 6 MeV protons in $MgB_2$ [7]. Defect concentration at the highest dose is around 0.1% in the range of the projectile with fairly bulk damage.

99.84% of the total energy loss is electronic and comes out to be ~1.5MeV/μ, considerably less than that required for generation of columnar tracks [16] and there is no sign of amorphous region as evident from XRD. Hence the defects generated are caused primarily by elastic nuclear collision giving rise to clusters of point defects of Mg and B.

### 3.1. Resistivity Analysis:

Though there has not been any significant change in $T_c$ after irradiation (except at the highest dose), we notice an appreciable increase in resistivity with irradiation. As already stated, in $MgB_2$, the grains are strongly coupled which are not disturbed even after irradiation, as noticed by inappreciable change in $\Delta T_c$ in contrast to HTSC cuprates. So, the intergranular damage is not primary cause for the increase in resistivty and the root of this increase is elsewhere.

$MgB_2$ is a strongly coupled phonon mediated superconductor. The decrease in resistivity linearly with temperature from 300K up to a certain point (~ 200 K) and then its deviation shows that resistivity can be explained from phonon scattering mechanism. We have fitted the experimental curve to Bloch-Grüneisen expression [17],

$$\rho(T) = \rho_0 + (m-1)\rho'\Theta\left(\frac{T}{\Theta}\right)^m \int_0^{\frac{\Theta}{T}} \frac{x^m \exp(x)}{(\exp(x)-1)^2} dx \qquad (1)$$

Here, $\rho_0$ is residual resistivity, $\rho'$ - temperature coefficient of resistivity and $\Theta$ the Debye temperature. $\rho_0$, $\rho'$ and $\Theta$ are the fitting parameters. Curve was fitted for m=3-5 by some groups [18,19], but m=5 seems more physical since in case of electron-phonon scattering, $\rho(T)$ varies as $T^5$ at low temperature. The fitted curves for the unirradiated sample and the one irradiated at the dose of $1 \times 10^{15}$ ions/cm$^2$ as representative cases are presented in fig 3. The values of $\rho_0$ and $\rho'$ for different samples are listed in Table 1. The increase in resistivity has contributions from $\rho_0$ and $\rho'$. The increase in $\rho_0$ can be related to the increase in defect concentration with irradiation. Decrease in $\rho_0$ at the lowest dose can be understood from annealing of the defects as already mentioned. Temperature coefficient of resistivity $\rho'$ is found to behave similarly to residual resistivity $\rho_0$ with irradiation. Debye temperature did not vary much with irradiation and was from 903K to 909K (variation is within the error range of the fit). The value of $\Theta$ obtained from various measurements of other groups vary from 800 K to 1050 K [4,18,20].

We have obtained the EPC constant $\lambda$ about 0.84 for the unirradiated sample using the experimentally obtained $T_c$ and fitted $\Theta$ value in the McMillan equation

$$T_c = \frac{\Theta}{1.45} \exp\left[\frac{-1.04(1+\lambda)}{\lambda - (1+0.62\lambda)\mu^*}\right] \qquad (2)$$

with the value of Coulomb pseudopotential $\mu^*$ taken as 0.1 [21]. $\lambda$ also has not changed significantly with irradiation. Our value of $\lambda$ is higher than 0.61-0.63 obtained from specific heat data [20,22], but close to that obtained from theoretical band structure calculations by Y. Kong et al. [17].

Observations of the resistivity measurements can be explained from the structure and multiband nature of $MgB_2$. As mentioned earlier, strong covalent σ-bonding within the B-B layer gives rise to 2D σ bands. The carriers of 2D σ bands are strongly coupled with the in-plane B $E_{2g}$ stretching modes, which gives rise to superconductivity [23,24]. Electron-phonon coupling constant along σ-band ($\lambda_\sigma$) governs $T_c$.

Contribution to conductivity is less in σ-band due to strong EPC. Rather, conductivity is large in π-band with low EPC constant and large value of density of states around Fermi surface (The contribution to $N(E_F)$ from π-bonding is 56% and that from σ-bonding is 44% [25]). There are both intralayer and interlayer π-bonding network within B atoms through $2p_z$ orbitals perpendicular to σ-bond. The interlayer π-bonding extends through Mg layers because of interaction of $Mg^{2+}$ ion with $2p_z$ orbitals of B [26]. Hence the 3-dimensional conductivity in $MgB_2$ occurs primarily due to the π band whereas in-plane B-B σ-bonding plays role in 2-Dimensional conductivity.

Particle irradiation causes vacancies in both B and Mg layers. Insignificant changes in λ, Θ and $T_c$ due to irradiation reflect strong B-B σ-bonding network. B and Mg vacancies bring forth damage in intralayer and inetrlayer π-bonding network respectively leading to substantial decrease in π-bonding $N(E_F)$. This is the primary cause of the increase in resistivity. As $\rho^,$ is inversely proportional to $N(E_F)$, decrease in $N(E_F)$ with irradiation causes an increase in $\rho^,$. This reflects on an increase in the temperature dependent part of resistivity, i.e. an increase in ($\rho_{300} - \rho_0$) with irradiation as shown in Table 1.

4. **Conclusion:**

We have carried out 160 MeV Neon ion irradiation at various doses up to $10^{15}$ ions/cm$^2$ on $MgB_2$ polycrystalline pellets. Resistivities of the samples were measured. $T_c$ of the material is insensitive to particle irradiation though resistivity increases. This is due to the presence of multiband nature of the Fermi surface. $T_c$ is governed by the coupling of in-plane B σ-band electrons with phonons from $E_{2g}$ vibration mode and is not changed much with irradiation up to the dose of $1 \times 10^{15}$ Neon/cm$^2$, as has been seen from Θ and λ extracted from fitting of the resistivity data to Bloch-Grüneisen expression. On the other hand, resistivity is controlled primarily by π-bonding network in intralayer and interlayer B. Irradiation causes damage in the π-bonding network through Mg and B vacancies giving rise to increase in resistivity. The decrease in resistivity at the lowest dose is due to the annealing of defects in the unirradiated samples. The insensitivity of $MgB_2$ towards the particle irradiation makes it a potential candidate for application in radiation environment.

**Acknowledgements:**

Authors like to thank Dr. B.Sinha, Director, VECC for his constant encouragement. One of the authors A. T. acknowledges CSIR, New Delhi, for financial support.

**References:**

[1] J. Nagamatsu, N. Nakagawa, T. Muranaka, Y. Zenitani, and J. Akimitsu, Nature 410 (2001) 63.

[2] D.C. Larbalestier, L.D. Cooley, M.O. Rikel, A.A. Polyanskii, J. Jiang, S. Patnaik, X.Y. Cai, D.F. Feldmann, A. Gurevich, A.A. Squitieri, M.T. Naus, C.B. Eom, E.E. Hellstorm, R.J. Cava, K.A. Regan, N. Rogdo, M.A. Hayward, T. He, J.S. Slusky, P. Khalifah, K. Inumaru and M. Hass, Nature 410 (2001) 186.

[3] S.L. Bud'ko, G. Lapertot, C. Petrovic, C.E. Cunningham, N. Anderson and P.C. Cannfield, Phys. Rev. Lett. 86 (2001) 1877.

[4] R.K. Kremer, B.J. Gibson and A. Ahn, cond-mat/0102432.

[5] J. Akimitsu and T. Muranaka, Physica C 388-389 (2003) 98

[6] F. Giubileo, D. Roditchev, W. Sacks, R. Lamy, D.X. Thanh, J. Klein, S. Miraglia, D. Fruchart, J. Marcus and Ph. Monod, Phys. Rev. Lett. 87 (2001) 177008.

[7] E. Mezzeti, D. Botta, R. Cherubini, A. Chiodoni, R. Gerbaldo, G. Ghigo, G. Giunchi, L. Gozzeline and B. Minetti, Physica C 372-376 (2002) 1277.

[8] A Gupta, H. Narayan, D. Astil, D. Kanjilal, C. Ferdeghini, M. Paranthaman and A.V. Narlikar, Supercond. Sci. Technol. 16 (2003) 951-955.

[9] G.K. Perkins, Y. Bugaslavsky, A.D. Caplin, J. Moore, T.J. Tate, R. Gwilliam, J. Jun, S.M. Kazakov, J. Karpiniski and L.F. Cohen, Supercond. Sci. Techhnol. 17 (2004) 232.

[10] S.K. Bandyopadhyay, P. Barat, Pintu Sen, A.K. Ghosh, A.N. Basu and B. Ghosh, Phys. Rev. B 58 (1998) 15135.

[11] O. Meyer, in: Studies of High Temperature Superconductors, edited by A.V. Narlikar, Vol. I. (Nova Sciences Publishers, New York, 1989), p 139 and references therein.

[12] A. Talapatra, S.K. Bandyopadhyay, Pintu Sen, A. Sarkar and P. Barat, Bull. Mat. Sc.[in Press].

[13] A.D. Marwick, G.J. Clark, D.S. Lee, R.B. Laidcwitz, G. Coleman and J.J. Como, Phys. Rev. Lett B 39 (1989) 9061.

[14] E.M. Schulson, J Nucl. Mater. 83 (1979) 239.

[15] J.F. Ziegler, www.SRIM.org

[16] G. Blatter, M.V. Feigel'man, V.B. Geshkenbein, A.I. Larkin and V.M. Vinokur, Rev. Mod. Phys. 66 (1994) 1125.

[17] Y. Kong, O.V. Dolgov, O. Jepsen and O.K. Anderson, Phys. Rev. B 64 (2001) 020501(R).

[18] A. Poddar, B. Bandyopadhyay, P. Mandal, D. Bhattacharya, P. Choudhury, U. Sinha and B. Ghosh, Physica C 390 (2003) 191.

[19] A. Bharathi, S.J. Balaselvi, S. Kalavathi, G.L.N. Reddy, V. Sankar Sastry, Y. Hariharan and T.S. Radhakrishnan, Physica C (2002) 211.

[20] F. Bouquet, R.A. Fisher, N.E. Philips, D.G. Hinks and J.D. Jorqensen, Phys. Rev. Lett. 87 (2001) 047001.

[21] I.I. Mazin and V.P. Antropov, Physica C 385 (2003) 49.

[22] Y.Wang, T.Plackowski and A.Junod, Physica C 355 (2001) 179.

[23] A.Y. Liu, I.I. Mazin and J. Kortus, Phys. Rev. Lett. 87 (2001) 087005.

[24] J.Kortus, I.I. Mazin, K.D. Belashchenko, V.P. Antropov and L.L.Boyer, Phys. Rev. Lett. 86 (2001) 4656.

[25] H.J. Choi, M.L. Cohen and S.G. Louie, Physica C 385 (2003) 66.

[26] E.Z. Kurmaev, I.I. Lyakhovskaya, J. Kortus, A. Moewes, N. Miyata, M. Demeter, M. Neumann, M. Yanagihara, M. Watanabe, T. Muranaka and J. Akimitsu, Phys. Rev. B 65 (2002) 134509.

**Figure captions**

**Figure 1.** X-ray diffraction patterns of MgB$_2$: (a) unirradiated ; (b) $1\times10^{13}$ ions/cm$^2$, (c) $1\times10^{14}$ ions/cm$^2$ and (d) $1\times10^{15}$ ions/cm$^2$.

**Figure 2.** Resistivity versus temperature of both unirradiated and irradiated samples of MgB$_2$.

**Figure 3.** Observed and fitted results of resistivity versus temperature for unirradiated samples and sample irradiated at $1\times10^{15}$ ions/cm$^2$ as representative.

**Table Caption.**

**Table 1.** T$_c$ and other resistivity parameters for irradiated and unirradiated samples

**Table 1.**

| Dose (ions/cm$^2$) | $T_c$ (K) | $\rho_0$ ($\mu\Omega$-cm) | $\rho_{300}$ ($\mu\Omega$-cm) | $\rho'$ ($\mu\Omega$-cm/K) | $\rho_{300} - \rho_0$ ($\mu\Omega$-cm) |
|---|---|---|---|---|---|
| Zero | 38.7 | 25.01 | 86.94 | 0.32 | 61.93 |
| 1X10$^{13}$ | 38.6 | 22.47 | 67.57 | 0.26 | 45.10 |
| 1X10$^{14}$ | 38.7 | 33.71 | 124.60 | 0.48 | 90.89 |
| 1X10$^{15}$ | 38.0 | 39.94 | 139.46 | 0.52 | 99.52 |

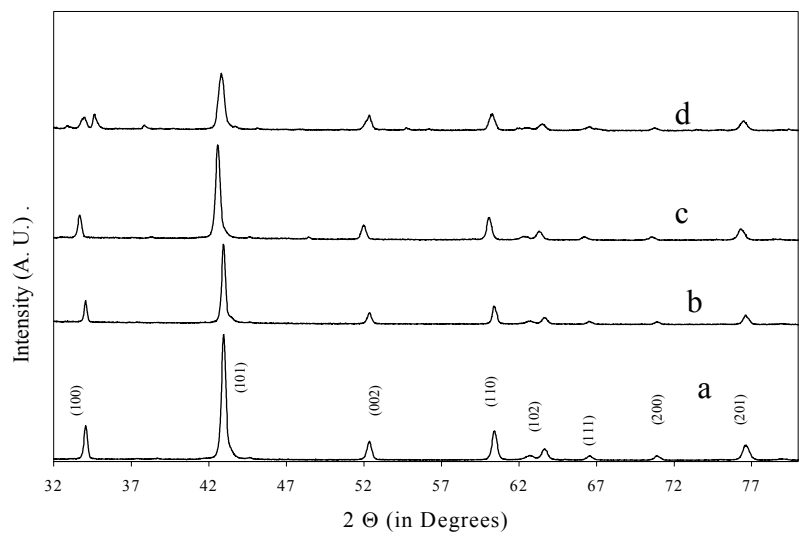

Figure 1.

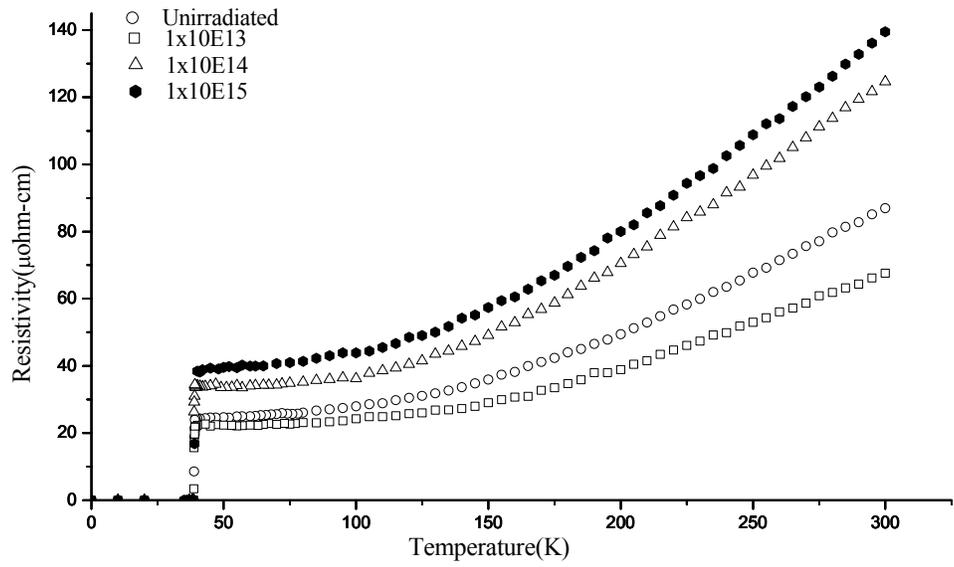

Figure 2.

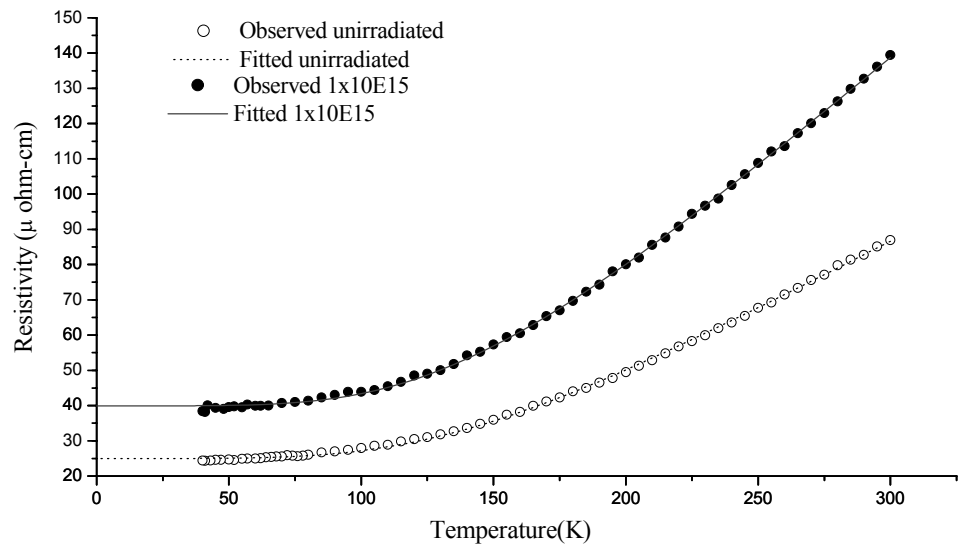

Figure 3.